\documentstyle[preprint,eqsecnum,aps,epsf,floats]{revtex}
\newdimen\pmboffset
\def\oldpmb#1{\setbox0=\hbox{#1}%
 \copy0\kern-\wd0 \kern\pmboffset\raise
 1.732\pmboffset\copy0\kern-\wd0 \kern\pmboffset\box0}
\def\pmb#1{\mathchoice{\oldpmb{$\displaystyle#1$}}{\oldpmb{$\textstyle#1$}}
      {\oldpmb{$\scriptstyle#1$}}{\oldpmb{$\scriptscriptstyle#1$}}}

\begin{document}
\draft
\preprint{CALT-68-2060}
\title{Color-Singlet $\pmb{\psi_Q}$ Production at $\pmb{e^+e^-}$ 
Colliders}
\author{Peter Cho and Adam K. Leibovich}
\address{
Lauritsen Laboratory\\
California Institute of Technology\\
Pasadena, CA 91125}
\date{\today}
\maketitle
\begin{abstract}
We calculate in closed form the complete ${\mathcal O}(\alpha_s^2)$ 
color-singlet differential cross section for 
$e^+e^- \to \gamma^* \to \psi_Q+X$ scattering.  The
cross section reduces at high energies to a heavy quark fragmentation form. 
We find that the energy scale at which the approximate fragmentation result
becomes reliable exceeds the $\psi_Q$ mass by more than an order of 
magnitude.  We also discuss the color-singlet model's predictions for 
direct $J/\psi$ angular and energy distributions at CLEO.
\end{abstract}
%\pacs{}
\newpage

% ==========================================================================
% Main body 
% ==========================================================================

% ==========================================================================
% Introduction
% ==========================================================================

%\narrowtext
\widetext

\section{Introduction}

During the past few years, there has been renewed interest in the study of
heavy quarkonium systems.  Much of the recent work on this subject has been
stimulated by large discrepancies between old predictions and new observations
of Psi and Upsilon production at several experimental facilities.  
Orders of magnitude disagreements between theory and data have seriously
undermined the conventional ``color-singlet model'' (CSM) picture of
quarkonia formation \cite{Chang,Berger,Kuhn,Guberina,Baier}.  
In this model, charmonia and bottomonia mesons
are presumed to exclusively originate from short distance processes that
create heavy quark-antiquark pairs in colorless configurations.  The quantum
numbers of pairs produced in high energy collisions on time scales 
short compared to $\Lambda_{QCD}$ are required to precisely match those of
the final state hadrons into which they nonperturbatively evolve.  
Although this
CSM picture is simple, it does not explain several gross features of recent
charmonia and bottomonia data collected at the Fermilab Tevatron
\cite{Psidata,Papadimitriou,Upsilondata}.  It
consequently must be abandoned as a complete theory.

A new framework for treating quarkonia systems called Nonrelativistic Quantum
Chromodynamics (NRQCD) has been developed within the past few years
\cite{Bodwin}.  This effective field
theory generalizes and improves upon the CSM in several regards.
It allows for short distance processes to create heavy 
quark-antiquark pairs in color-octet configurations which can hadronize over
much longer length scales into colorless final state quarkonia.
Calculations which include this color-octet mechanism
appear to successfully describe Tevatron measurements
\cite{BraatenFleming,CGMP,ChoLeibov1,ChoLeibov2}.
But in order to 
establish the validity of this new paradigm, it is 
necessary to consider quarkonia production in other experimental 
situations.

Braaten and Chen have suggested that a clean signature of the color-octet
mechanism may be observable in $\psi_Q$ production at electron-positron
colliders \cite{BraatenChen}.
These authors have noted that the angular distribution of colored 
$Q\bar{Q}$ pairs near the endpoint region may qualitatively differ from 
that of their colorless counterparts.  If this effect
could be observed, it would support the color-octet
production picture.  It might also permit an independent 
determination of the numerical values for certain NRQCD matrix elements.

Before the search for color-octet quarkonia production in $e^+e^-$ 
annihilation can begin, one must first know the precise CSM prediction.
Within the NRQCD framework, the color-singlet cross section is also expected
to be quite accurate for all energies except near the endpoint region
\cite{BraatenChen}.
In this note, we therefore build upon previous studies reported in the
literature \cite{Kuhn1,Kuhn2,Driesen,Clavelli} and calculate 
the complete ${\mathcal O}(\alpha_s^2)$ color-singlet cross
section for $e^+e^- \to \psi_Q + X$ scattering.  We 
examine the contribution to $\psi_Q$ production from the short distance
modes 
$e^+e^- \to Q\bar{Q}[{}^3S_1^{(1)}]+g+g$ and 
$e^+e^- \to Q\bar{Q}[{}^3S_1^{(1)}]+Q+\bar{Q}$,\footnote{We indicate the 
angular momentum and color-singlet
quantum numbers of the $Q\bar{Q}$ pair which hadronizes into the final
state $\psi_Q$ meson inside square brackets.}
and we derive a closed form expression for the differential cross section.
We then discuss the implications of the CSM result for direct $J/\psi$ 
observations at CLEO.  Finally, we compare heavy quark fragmentation 
predictions with the color-singlet cross section and determine the energy 
scale at which fragmentation approximations become reliable.
% ==========================================================================
% General Diff. x-section discussion
% ==========================================================================

\section{inclusive angular distributions in electron-positron collisions}

It is useful to note some general features of inclusive, unpolarized 
$\psi_Q$ production in $e^+e^-$ annihilation.  Unitarity, parity and angular
momentum considerations restrict the form of the differential cross section
expression
\begin{equation}\label{generalcrosssection}
\frac{d^2\sigma}{dE_3d\cos\theta_3}
\left(e^+(p_1)e^-(p_2) \to \gamma^* \to \psi_Q(p_3) + X\right) =
S(E_3)\left[1 + \alpha(E_3)\cos^2\theta_3\right].
\end{equation}
In particular, the allowed range for the angular coefficient function is
constrained to lie within the interval $-1 \leq \alpha(E_3) \leq 1$.  We 
sketch a derivation of this result below.

It is instructive to consider the subprocess 
$\gamma^*(P) \to \psi_Q(p_3) + X(P-p_3)$ 
where the intermediate photon is either longitudinally or transversely aligned.
The squared amplitude for this decay
\begin{equation}\label{gammaAX}
|{\mathcal A}|^2 = 
\sum_{\lambda} \varepsilon_\mu(P;\lambda) \varepsilon_\nu(P;\lambda)^* 
F^{\mu\nu}\label{gammaAXb}
\end{equation}
involves a form factor $F^{\mu\nu}$ 
which can be decomposed in terms of tensors that respect parity and gauge 
invariance:
\begin{equation}\label{formfactors}
F^{\mu\nu} = -F_1\left(g^{\mu\nu} - \frac{P^\mu P^\nu}{P^2}\right)
+ \frac{F_2}{P^2}\left(p_3^\mu - \frac{P\cdot p_3}{P^2}P^\mu\right)
\left(p_3^\nu - \frac{P\cdot p_3}{P^2}P^\nu\right).
\end{equation}
Working in the $\gamma^*$ rest frame where the $\psi_Q$ four-momentum looks
like 
$p_3 = (E_3,\vec{p}_3) 
= (E_3,|\vec{p}_3|\sin\theta\cos\phi,|\vec{p}_3|\sin\theta\sin\phi,
|\vec{p}_3|\cos\theta)$,
we find that the squared decay amplitude for a longitudinally polarized
virtual photon reduces to
\begin{mathletters}\label{amplong}
\begin{equation}\label{amplongsq}
|{\mathcal A}|^2_L =  F_1[1+\alpha_L\cos^2\theta],
\end{equation}
with
\begin{equation}\label{alphalong}
\alpha_L = \frac{|\vec{p}_3|^2}{P^2}\, \frac{F_2}{F_1}.
\end{equation}\end{mathletters}%
For a transverse 
$\gamma^*$, the squared amplitude takes the form
\begin{mathletters}\label{amptran}
\begin{equation}\label{amptransq}
|{\mathcal A}|^2_T = 
\left(2 F_1 + \frac{|\vec{p}_3|^2}{P^2}F_2\right)[1+\alpha_T\cos^2\theta],
\end{equation}
where
\begin{equation}\label{alphatran}
\alpha_T = -\frac{|\vec{p}_3|^2 F_2}{2P^2F_1 + |\vec{p}_3|^2 F_2}
= -\frac{\alpha_L}{2+\alpha_L}.
\end{equation}\end{mathletters}

\noindent Since both $|{\mathcal A}|^2_L$ and
$|{\mathcal A}|^2_T$ are nonnegative, eqs.~(\ref{amplong}) and (\ref{amptran})
imply $\alpha_L \geq -1$ and $-1~\leq~\alpha_T~\leq~1$.  

Helicity conservation requires the intermediate photon in 
$e^+e^- \to \gamma^* \to \psi_Q+X$ to be transversely aligned relative to the
beam axis in the $m_e = 0$ limit.  The $\psi_Q$ meson's angular
distribution is therefore significantly restricted by simple symmetry 
considerations.  In fact, the inclusive angular distribution of {\it any}
unpolarized particle which is produced in electron-positron colliders
operating well below the $Z$-pole goes as $1 + \alpha_T \cos^2\theta$ with
$-1 \leq \alpha_T \leq 1$.  So while observation of a pure $\sin^2\theta$ 
distribution for a lepton or hadron at a collider like CLEO is possible,
a pure $\cos^2\theta$ distribution is not.  As we shall see, all color-singlet
$\psi_Q$ predictions are consistent with these general considerations.

% ==========================================================================
% Calculation of cross section
% ==========================================================================

\section{Color-singlet $\pmb{\psi_Q}$ production}

The simplest parton level process which mediates color-singlet production
of $J^{PC}=1^{--}$ quarkonia is given by
$e^+e^-\to Q\bar{Q}[{}^3S_1^{(1)}]+g+g$.  Color, parity and charge conjugation
conservation require two gluons to appear in the final state along with the
colorless $Q\bar{Q}[{}^3S_1^{(1)}]$ pair.  This channel consequently
 contributes to the
$\psi_Q$ cross section starting at ${\mathcal O}(\alpha_s^2)$.  Color singlet 
production also proceeds at the same order in perturbative QCD through the
mode $e^+e^-\to Q\bar{Q}[{}^3S_1^{(1)}]+Q+\bar{Q}$.  These two distinct
reactions have been considered separately in the literature
\cite{Kuhn1,Kuhn2,Driesen,Clavelli}.  We will reexamine their joint impact
upon Psi and Upsilon production and derive a closed form analytic expression
for $d^2\sigma/dE_3\,d\cos\!\theta_3$.  We can then compare the relative
magnitudes of the gluon and quark processes as a function of center-of-mass
energy~$\sqrt{S}$.

The leading order diagrams which mediate 
$e^+(p_1)\,e^-(p_2)\to Q\bar{Q}[{}^3S_1^{(1)}](p_3)+g(p_4)+g(p_5)$ and 
$e^+(p_1)\,e^-(p_2)\to Q\bar{Q}[{}^3S_1^{(1)}](p_3)+Q(p_4)+\bar{Q}(p_5)$
scattering are illustrated in Figs.~1 and~2.  
The hard collisions pictured in the figures form on short time scales a heavy 
quark and antiquark
which fly out from the primary interaction point in nearly parallel directions
and almost on-shell.  The $Q\bar{Q}$ pair then evolves over a much longer time
interval into a physical $\psi_Q$ bound state.  Working within the NRQCD 
framework and using computational methods discussed in 
Refs.~\cite{Guberina,ChoLeibov1,ChoLeibov2}, one can straightforwardly 
calculate the amplitudes for these processes.  Their squares factorize into 
products of short distance coefficient functions and long 
distance NRQCD matrix elements.

\begin{figure}[ht]
\centerline{\epsfysize=6truecm \epsfbox{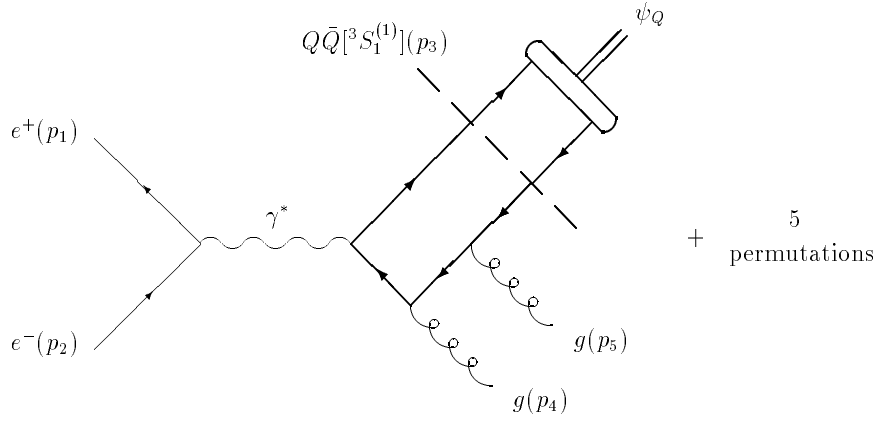}}
\caption[1]{Leading order Feynman graphs which mediate
$e^+e^- \to \gamma^* \to Q\bar{Q}[{}^3S_1^{(1)}] + g + g$$\to \psi_Q + X$ 
scattering.}
\end{figure}

\begin{figure}[ht]
\centerline{\epsfysize=11truecm \epsfbox{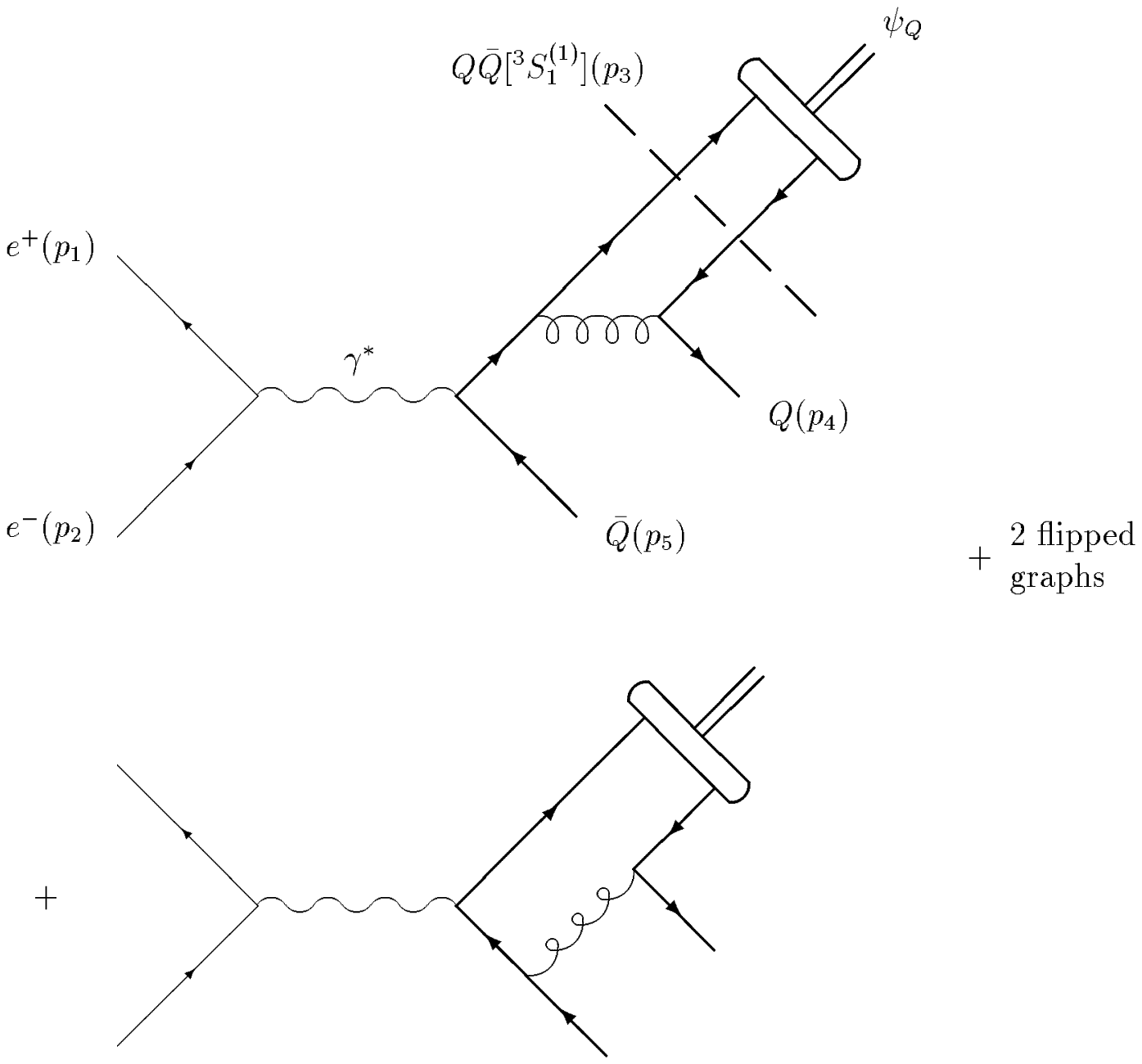}}
\caption[2]{Leading order Feynman graphs which mediate
$e^+e^- \to \gamma^* \to Q\bar{Q}[{}^3S_1^{(1)}]+Q+\bar{Q}$\\
$\to \psi_Q + X$ scattering.}
\end{figure}

Integrating the squared amplitudes over the three 
particle phase space factor
\begin{equation}\label{completephasespace}
d\Phi_3 = (2\pi)^4\delta^4(p_1+p_2-p_3-p_4-p_5) 
\prod_{i=3}^5\frac{d^3p_i}{(2\pi)^3 2E_i}
\end{equation}
is somewhat involved.  As a simplifying measure, we rescale all 
dimensionful quantities relative to
the beam energy $E$ and work with the dimensionless variables
$z_i = E_i/E$, $\vec{q}_i~=~\vec{p}_i/E$, $x_i = \cos\theta_i$ and 
$\delta = 2 M_Q/E$.
The phase space factor for the reaction with gluons in the final state can 
then be reduced to the form
\begin{equation}\label{simplephasespace}
d\Phi_3 = \frac{(2\pi)}{8}^{-4}\,E^2\,\frac{\,dz_3\,dx_3\,dz_{-}\,dw}
	{\sqrt{(1-K^2)(1-x_3^2)-w^2}}
\end{equation}
where 
\begin{mathletters}\label{phasespacevariables}
\begin{eqnarray}
z_- &=& z_4 - z_5\\
|\vec{q}_{-}| &=& \sqrt{4 - 4z_3 + \delta^2 + z_{-}^2}\\
|\vec{q}_3| &=& \sqrt{z_3^2 - \delta^2}\\
K &=& \frac{z_{-}(2-z_3)}{|\vec{q}_{-}||\vec{q}_3|}\\
w &=& x_{-} + K x_3.
\end{eqnarray}
\end{mathletters}

\noindent The same result holds for the quark process with the simple
alteration $|\vec{q}_{-}| = \sqrt{4 - 4z_3 + z_{-}^2}$.

The available phase space volume clearly depends upon the masses of the
final state bodies.  For the $e^+e^- \to Q\bar{Q}[{}^3S_1^{(1)}]+g+g$ channel,
the limits of integration for the remaining energy and angular variables in 
eq.~(\ref{simplephasespace}) are given by 
\begin{mathletters}\label{phasespacelimits}
\begin{eqnarray}
\delta \leq &z_3& \leq 1 + \frac{\delta^2}{4}\\
-1 \leq &x_3& \leq 1\\
-\sqrt{z_3^2 - \delta^2} \leq &z_{-}& \leq \sqrt{z_3^2 - \delta^2}\\
-\sqrt{(1-K^2)(1-x_3^2)} \leq &w& \leq \sqrt{(1-K^2)(1-x_3^2)}.
\end{eqnarray}
\end{mathletters}

\noindent The corresponding limits for the 
$e^+e^-\to Q\bar{Q}[{}^3S_1^{(1)}]+Q+\bar{Q}$ mode
\begin{mathletters}\label{heavyphasespacelimits}
\begin{eqnarray}
\delta \leq &z_3& \leq 1\\
-1 \leq &x_3& \leq 1\\
-\sqrt{\frac{(4 - 4z_3)(z_3^2 - \delta^2)}{4 - 4z_3+\delta^2}}
        \leq &z_{-}& \leq
        \sqrt{\frac{(4 - 4z_3)(z_3^2 - \delta^2)}{4 - 4z_3+\delta^2}}\\
-\sqrt{(1-K^2)(1-x_3^2)} \leq &w& \leq \sqrt{(1-K^2)(1-x_3^2)}
\end{eqnarray}
\end{mathletters}
are more tight due to the additional heavy quark and antiquark in the
final state.

After inserting the averaged squared amplitudes and reduced phase space factors
into the formula
\begin{equation}\label{crosssection}
d\sigma = \frac{1}{8E^2}\overline{\sum}|{\mathcal A}|^2d\Phi_3,
\end{equation}
we can analytically integrate over $w$ and $z_-$ and obtain differential
expressions of the form (\ref{generalcrosssection}).  We display the resulting
$S(z_3)$ and $\alpha(z_3)$ functions for the 
$e^+e^-~\to~Q\bar{Q}[{}^3S_1^{(1)}]+g+g$ and 
$e^+e^-~\to~Q\bar{Q}[{}^3S_1^{(1)}]+Q+\bar{Q}$ processes in the appendix.
As a check, one can verify that $|\alpha_{\mathrm gluon}|$ and
$|\alpha_{\mathrm quark}|$ do not exceed unity within their allowed $z_3$ 
ranges as required by the general constraints discussed in section II.
The total ${\mathcal O}(\alpha_s^2)$ angular coefficient function
\begin{equation}\label{alphasinglet}
\alpha_{\mathrm total} = 
\frac{S_{\mathrm gluon}\,\alpha_{\mathrm gluon} + S_{\mathrm quark}\,
\alpha_{\mathrm quark}}{S_{\mathrm gluon} + S_{\mathrm quark}}
\end{equation}
also respects the bound $-1 \leq \alpha_{\mathrm total} \leq 1$.

Another important check can be performed by considering the high energy 
behavior of the $S(z_3)$ and $\alpha(z_3)$ functions.  In the $z_3 \gg \delta$ 
limit, the color-singlet cross section reduces to
\begin{eqnarray}\label{frag}
\frac{d^2\sigma}{dz_3d\cos\theta_3}(e^+e^- \to \psi_Q + X) &=&
\frac{4 \pi}{243}
\frac{(\alpha_s\alpha_{\scriptscriptstyle EM} {\mathcal Q}_Q)^2}{m_Q^3 E^2}
\left <0|{\mathcal O}^{\psi_Q}_1(\,{}^3S_1)|0\right>
\left[1 + \cos^2\theta_3\right]\\
&&\times 
\frac{z_3(1-z_3)^2(16 - 32 z_3 + 72 z_3^2 - 32 z_3^3 + 5 z_3^4)}
{(2-z_3)^6}.\nonumber
\end{eqnarray}
After integrating over $\cos\theta_3$ and recalling the relation 
$\left <0|{\mathcal O}^{\psi_Q}_1(\,{}^3S_1)|0\right> = 
9|{\mathcal R}(0)|^2/2\pi$ 
between the color-singlet NRQCD matrix element and the $\psi_Q$ wavefunction 
at the origin \cite{Bodwin}, we can write the $\psi_Q$ energy distribution as
\begin{equation}\label{fragform}
\frac{d\sigma}{dz_3}(e^+e^- \to \psi_Q + X) = 
2 \sigma(e^+e^- \to Q\bar{Q})\times {\mathcal D}_{Q \to \psi_Q}(z_3)
\end{equation}
where ${\mathcal D}_{Q \to \psi_Q}(z_3)$ denotes the heavy quark 
fragmentation function calculated in Ref.~\cite{BCY}.  The complete
${\mathcal O}(\alpha_s^2)$ color-singlet cross section thus correctly 
reproduces known fragmentation results at high energies.

\section{Direct $\pmb{J/\psi}$ production at CLEO}

$J/\psi$ production is currently under study at CLEO\cite{CLEO1,CLEO2}.
Charmonia observed at this $e^+e^-$ facility mainly come from $B$ meson
decays.  However, a clean sample of $\psi$'s originating from continuum 
production can be obtained by imposing a lower momentum cut on their dilepton
decay products.  Various 
characteristics of the resulting direct $J/\psi$ data sample can then be 
compared with predictions based upon color-singlet and color-octet production
mechanisms.  Such experimental investigations are underway\cite{Wolf}.

The angular distribution of direct $J/\psi$ mesons represents one
observable which can be measured at CLEO.  In Fig.~3, we plot the CSM 
prediction for the angular coefficient function $\alpha$.
The results displayed in the figure are based upon the 
input parameter values $E = 5.29$ GeV, $m_c = 1.48$ GeV, $\alpha_s(2m_c)=0.28$,
$\alpha_{\scriptscriptstyle EM}(2m_c)=0.0075$, ${\mathcal Q}_c = 2/3$ and 
$\left <0|{\mathcal O}^{J/\psi}_1(\,{}^3S_1)|0\right> = 1.2\ {\mathrm GeV}^3$. 
The dashed curve illustrates the function $\alpha_{\mathrm gluon}$ associated 
with $e^+e^- \to c\bar{c}[{}^3S_1^{(1)}]+g+g$ scattering.  
\begin{figure}[ht]
\centerline{\epsfysize=12truecm \epsfbox{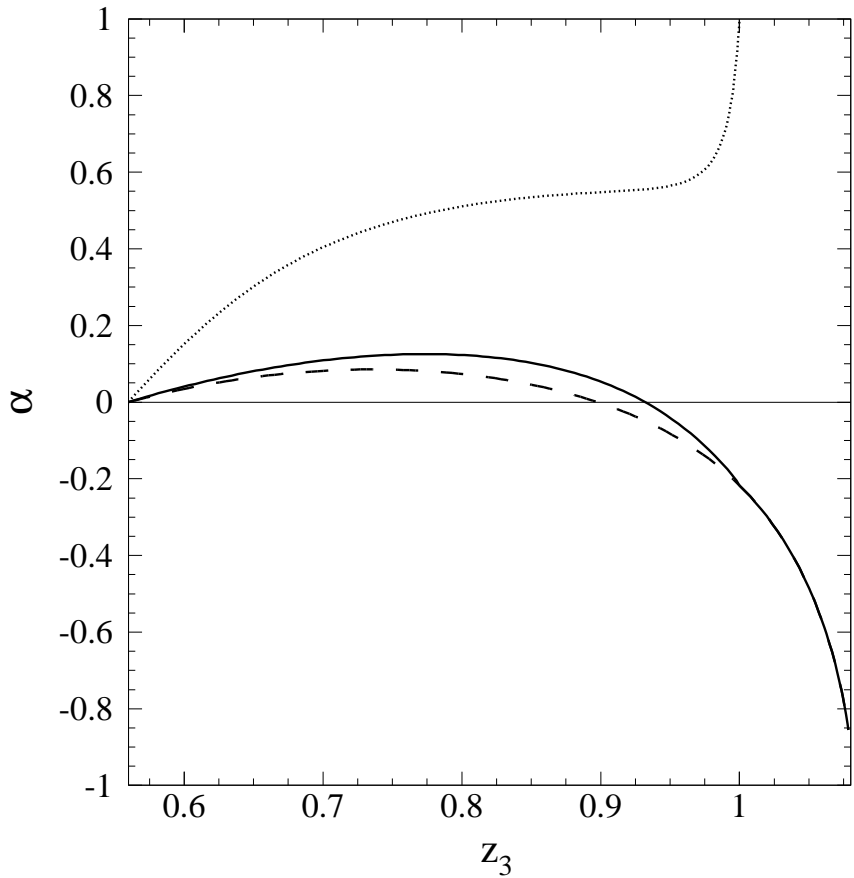}}
\caption[3]{Angular coefficient functions $\alpha_{\mathrm gluon}$
(dashed line),
$\alpha_{\mathrm quark}$ (dotted line) and 
$\alpha_{\mathrm total}$ (solid line) plotted against dimensionless energy
variable $z_3$.}
\end{figure}
The shape of this curve
agrees with numerical results of Driesen {\it et al.} reported in 
Ref.~\cite{Driesen}.  The dotted line in Fig.~3 depicts the function 
$\alpha_{\mathrm quark}$ originating
from the $e^+e^- \to c\bar{c}[{}^3S_1^{(1)}]+c+\bar{c}$ mode.  The shape of
$\alpha_{\mathrm quark}$ is clearly quite different than that of 
$\alpha_{\mathrm gluon}$.  But since $S_{\mathrm quark}$ is substantially
smaller than $S_{\mathrm gluon}$ at CLEO energies, it has only a small impact
upon the total color-singlet function $\alpha_{\mathrm total}$ which is 
represented by the solid curve in Fig.~3.  It is important to note that
$\alpha_{\mathrm total}$ is predicted within the CSM to be negative at the 
largest allowed values for $z_3$.  On the other hand, color-octet effects may 
render $\alpha_{\mathrm total}$ positive in the endpoint region 
\cite{BraatenChen}.  The angular distribution of the most energetic $J/\psi$'s
at CLEO can therefore provide a valuable test of the color-octet mechanism.

The energy distribution of direct  $J/\psi$'s is another quantity
which can be used to probe theories of quarkonia production.  
In Fig.~4, we display the separate contributions to $d\sigma/dz_3$ from the
$e^+e^- \to c\bar{c}[{}^3S_1^{(1)}]+g+g$ and
$e^+e^- \to c\bar{c}[{}^3S_1^{(1)}]+c+\bar{c}$ channels along with the total
CSM prediction.  
\begin{figure}[ht]
\centerline{\epsfysize=12truecm \epsfbox{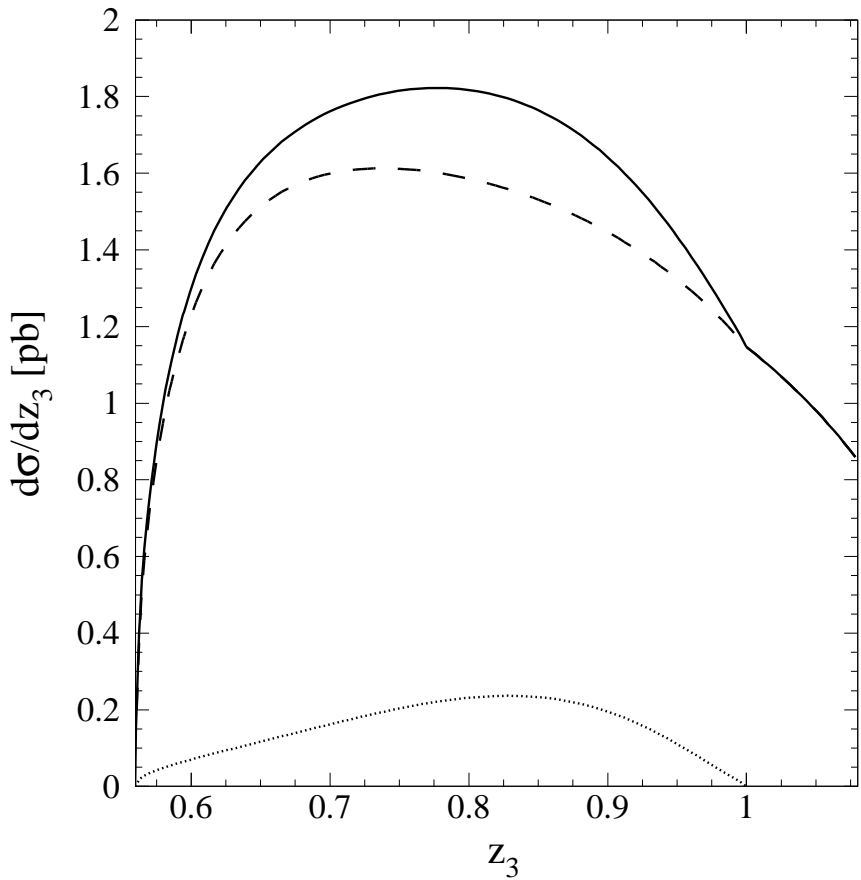}}
\caption[4]{Contributions to $d\sigma/dz_3$ from the gluon (dashed line) and
quark (dotted line) modes plotted against $z_3$.  The CSM prediction for the
total direct $J/\psi$ energy distribution is represented by the solid curve.}
\end{figure}
The sensitivity of this energy observable to the charm mode
is more pronounced than that of the angular coefficient function.
The areas underneath the dashed, dotted and solid curves respectively
equal $0.74$ pb, $0.07$ pb and $0.81$ pb.  The quark process thus contributes
at the 10\% level to direct $J/\psi$ production at CLEO.

The $e^+e^- \to c\bar{c}[{}^3S_1^{(1)}]+c+\bar{c}$ mode is significantly phase
space suppressed compared to $e^+e^- \to c\bar{c}[{}^3S_1^{(1)}]+g+g$ at CLEO
energies.  As a result, its impact upon charmonia observables is 
minor.  However, it is interesting to examine the relative importance of these
two color-singlet channels as a function of center-of-mass energy.  We plot in
Fig.\ 5 the modes' separate contributions to the integrated $J/\psi$ cross
section along with their sum versus $\sqrt{S} = 2E$.  We also display the
integral of the charm quark fragmentation approximation (\ref{frag}).
At low energies, the charm quark mode is negligible
compared to its gluon counterpart.  
\begin{figure}[ht]
\centerline{\epsfysize=12truecm \epsfbox{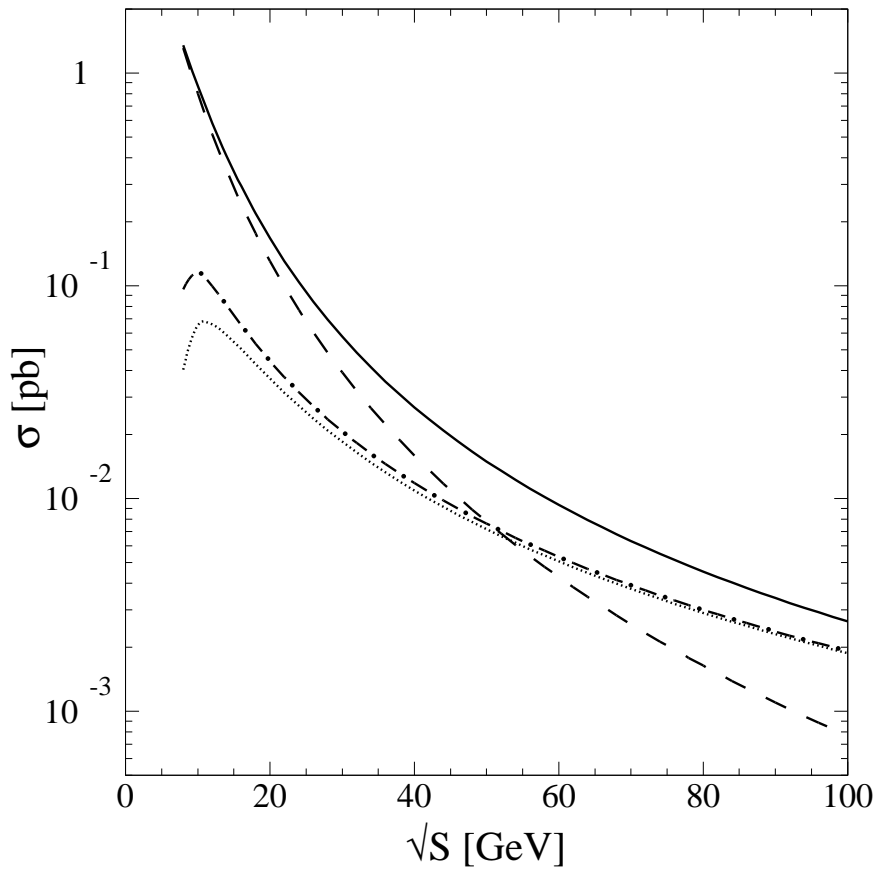}}
\caption[5]{Integrated cross sections for the gluon (dashed line) and
charm quark (dotted line) modes plotted  as a function of 
$\sqrt{S}$.  The sum of the two is shown by the solid curve.
The approximate charm quark fragmentation cross section is depicted by the 
dot-dashed curve.}
\end{figure}
At larger values of $\sqrt{S}$, it becomes 
relatively more important.  Finally, in the charm fragmentation limit 
$\sqrt{S} \gg m_c$, the quark mode dominates.  

As can be seen in Fig.~5, the
charm quark fragmentation curve rapidly asymptotes to the 
$e^+e^- \to c\bar{c}[{}^3S_1^{(1)}]+c+\bar{c}$ cross section.  
But it is important to note that the crossover point at which
the rates for the charm and gluon modes become equal occurs around
$\sqrt{S} \simeq 50$ GeV.  Consequently, the fragmentation approximation does
{\it not} accurately reflect the total color-singlet cross section until
$\sqrt{S}$ exceeds $2m_c$ by more than an order of magnitude.  This result for
$J/\psi$ production at lepton colliders is quite different than that for 
hadron accelerators.  Previous investigations have found that fragmentation
approximations are reasonably trustworthy for production of $\psi$'s
at the Tevatron with $p_\perp \gtrsim 10$ GeV \cite{ChoLeibov1,ChoLeibov2}.  
The moral we thus draw from this study of charmonia at CLEO 
is that the validity of fragmentation predictions must be 
carefully checked on a case-by-case basis.

\acknowledgments

We thank David Politzer for helpful discussions.  The work of P.C. was 
supported in part by a DuBridge Fellowship and by the Department of Energy 
under Grant No. DE-FG03-92-ER40701.  The work of A.L. was supported in 
part by the DOE under Grant No. DE-FG03-92-ER40701.

\eject
\appendix
\section*{}

We list here the color-singlet functions $S$ and 
$\alpha$ which enter into the
differential cross section (\ref{generalcrosssection}) at leading order in both
the perturbative QCD and NRQCD velocity expansions.  The contributions from 
the $e^+e^- \to Q\bar{Q}[{}^3S_1^{(1)}]+g+g$ and 
$e^+e^- \to Q\bar{Q}[{}^3S_1^{(1)}]+Q+\bar{Q}$ channels are separately 
displayed.

\vskip 1cm
\noindent \underline{$e^+e^- \to Q\bar{Q}[{}^3S_1^{(1)}]+g+g$ mode:}
\begin{mathletters}\label{boxsandalpha}
\begin{eqnarray}\label{boxs}
S_{\mathrm gluon}&=& \frac{\pi}{216}
\frac{(\alpha_s\,\alpha_{\scriptscriptstyle EM}\,
{\mathcal Q}_Q)^2}{\delta\,E^5} 
\frac{\left <0|{\mathcal O}^{\psi_Q}_1(\,{}^3S_1)|0\right>}
{(z_3 - 2)^2\,(2 z_3 - \delta^2)^3\,(z_3^2 - \delta^2)}\nonumber\\
&\times& \left\{\frac{}{}
4 \left[-\delta^2(4+\delta^2)(48 + 48 \delta^2 + 13 \delta^4)
+ 32 \delta^2(4 + \delta^2)(4 + 3 \delta^2)z_3\right.\right.
\nonumber\\
&&\qquad +\,8(32 - 56 \delta^2 - 24 \delta^4 + \delta^6)z_3^2 
-16(32 + 4 \delta^2 + 3 \delta^4)z_3^3\nonumber\\
&&\qquad\left.+\,112 (4 + \delta^2)z_3^4 - 128 z_3^5\right]
(2z_3-\delta^2)\sqrt{z_3^2 - \delta^2}\\
&&\quad +\,\left[\delta^4(4-\delta^2)(48 + 96 \delta^2 + 13 \delta^4) - 
32\delta^4(28-3 \delta^2 - 3 \delta^4)z_3\right.\nonumber\\
&&\qquad +\,8\delta^2(16 - 40 \delta^2 - 27 \delta^4 + \delta^6)z_3^2 +
16\delta^2(56 + 14\delta^2 - 3 \delta^4)z_3^3\nonumber\\
&&\left.\left.\qquad-\,16(4-\delta^2)(4 + 5\delta^2)z_3^4\right]
(4 z_3 - 4 - \delta^2)\ln
\frac{2z_3 - \delta^2 + 2\sqrt{z_3^2 - \delta^2}}
{2z_3 - \delta^2 - 2\sqrt{z_3^2 - \delta^2}}\right\}\nonumber
\end{eqnarray}
\begin{eqnarray}\label{boxalpha}
\alpha_{\mathrm gluon}(z_3) &=& \frac{\pi}{216}
\frac{(\alpha_s\,\alpha_{\scriptscriptstyle EM}\,
{\mathcal Q}_Q)^2}{\delta\,E^5} 
\frac{\left <0|{\mathcal O}^{\psi_Q}_1(\,{}^3S_1)|0\right>}
{(z_3 - 2)^2\,(2 z_3 - \delta^2)^3\,(z_3^2 - \delta^2)}\nonumber\\
&\times& \left\{\frac{}{}
4 \left[\delta^2(64 + 80 \delta^2 + 76 \delta^4 + 7 \delta^6)
- 96 \delta^4(4 + \delta^2)z_3\right.\right.
\nonumber\\
&&\qquad-\,
8(32 - 40 \delta^2 - 44 \delta^4 - \delta^6)z_3^2 
-16\delta^2(28 + 3 \delta^2)z_3^3\nonumber\\
&&\qquad\left.+\,16 (20 + 7 \delta^2)z_3^4 - 128 z_3^5\right]
(2z_3-\delta^2)\sqrt{z_3^2 - \delta^2}\\
&&\quad -\,\left[\delta^4(4-\delta^2)(4+\delta^2)(4 + 7 \delta^2) - 
32\delta^4(1-\delta^2)(4+3 \delta^2)z_3\right.\nonumber\\
&&\qquad -\,8\delta^2(16 + 40 \delta^2 + 57 \delta^4 + \delta^6)z_3^2 +
16\delta^2(8 + 58\delta^2 + 3 \delta^4)z_3^3\nonumber\\
&&\left.\left.\qquad+\,16(16-32\delta^2 - 5\delta^4)z_3^4\right]
(4 z_3 - 4 - \delta^2)\ln
\frac{2z_3 - \delta^2 + 2\sqrt{z_3^2 - \delta^2}}
{2z_3 - \delta^2 - 2\sqrt{z_3^2 - \delta^2}}\right\}\times 
\frac{1}{S_{\mathrm gluon}(z_3)}\nonumber
\end{eqnarray}
\end{mathletters}
\vskip 1cm

\noindent \underline{$e^+e^- \to Q\bar{Q}[{}^3S_1^{(1)}]+Q+\bar{Q}$ mode:}
\begin{mathletters}\label{nonboxsandalpha}
\begin{eqnarray}\label{nonboxs}
S_{\mathrm quark}(z_3) &=& \frac{\pi}{3888}
\frac{(\alpha_s\,\alpha_{\scriptscriptstyle EM}\,
{\mathcal Q}_Q)^2}{\delta^3\,E^5} 
\frac{\left <0|{\mathcal O}^{\psi_Q}_1(\,{}^3S_1)|0\right>}
{z_3^3\,(z_3 - 2)^6\,(z_3^2 - \delta^2)}\nonumber\\
&\times& \left\{
4 z_3\sqrt{\frac{(1 - z_3)(z_3^2-\delta^2)}
{4 + \delta^2 - 4 z_3}}
\left[-32\delta^4(4+\delta^2)(48 + 22 \delta^2 + 3 \delta^4)\right.\right.
\nonumber\\
&&\qquad+\,32 \delta^4(768 + 400 \delta^2 + 66 \delta^4 + 3 \delta^6)z_3
\nonumber\\
&&\qquad-\,16 \delta^2(384 + 1920 \delta^2 + 556 \delta^4 + 29 \delta^6 - 
2\delta^8)z_3^2
\nonumber\\
&&\qquad+\,
8\delta^2(1792 + 128 \delta^2 - 568 \delta^4-80\delta^6 - \delta^8)
z_3^3\nonumber\\
&&\qquad+\,2 (2048 - 11008 \delta^2 + 10752 \delta^4 + 3176 \delta^6
 + 98 \delta^8 + 3 \delta^{10} )z_3^4\nonumber\\
&&\qquad-\,4(4096 - 7808 \delta^2 + 3424 \delta^4 + 600 \delta^6 + 17 \delta^8)
z_3^5\nonumber\\
&&\qquad+\,(38912 - 20608 \delta^2 + 4544 \delta^4 + 508 \delta^6 - 3 \delta^8)
z_3^6\nonumber\\
&&\qquad-\,4(13312 - 800 \delta^2 + 120\delta^4 - 3\delta^6)z_3^7 +
8(4512 - 20 \delta^2 - 15 \delta^4)z_3^8\nonumber\\
&&\qquad-\,\left.32(336-\delta^2)z_3^9 + 1280 z_3^{10}\right]\\
&&\quad -\,\left[8\delta^4(48+22\delta^2 + 3 \delta^4) - 
32\delta^4(24+5 \delta^2)z_3\right.\nonumber\\
&&\qquad -\,2\delta^2(448 + 16 \delta^2 + 8 \delta^4 - 3 \delta^6)z_3^2 +
16\delta^2 (56 - 10\delta^2 - 5 \delta^4)z_3^3\nonumber\\
&&\qquad +\,\delta^2(1152 + 272 \delta^2 - 3 \delta^4)z_3^4 +
8(32 - 92 \delta^2 + 5 \delta^4)z_3^5\nonumber\\
&&\left.\qquad-\,56(16+\delta^2)z_3^6 + 512 z_3^7\right]\nonumber\\
&&\left.\qquad\times\delta^2(z_3^2 - 2)^4\ln
\frac{z_3\,\sqrt{4 + \delta^2 - 4 z_3} + 2\sqrt{(1-z_3)(z_3^2 - \delta^2)}}
{z_3\,\sqrt{4 + \delta^2 - 4 z_3} - 2\sqrt{(1-z_3)(z_3^2 - \delta^2)}}
\right\}\nonumber
\end{eqnarray}

\begin{eqnarray}\label{nonboxalpha}
\alpha_{\mathrm quark}(z_3) &=& \frac{\pi}{3888}
\frac{(\alpha_s\,\alpha_{\scriptscriptstyle EM}\,
{\mathcal Q}_Q)^2}{\delta^3\,E^5} 
\frac{\left <0|{\mathcal O}^{\psi_Q}_1(\,{}^3S_1)|0\right>}
{z_3^3\,(z_3 - 2)^6\,(z_3^2 - \delta^2)}\nonumber\\
&\times& \left\{
4 z_3 \sqrt{\frac{(1 - z_3)(z_3^2-\delta^2)}
{4 + \delta^2 - 4 z_3}}
\left[32\delta^4(4+\delta^2)(16 + 2 \delta^2 + 3 \delta^4)\right.\right.
\nonumber\\
&&\qquad-\, 32 \delta^4(256 + 48 \delta^2 + 22 \delta^4 + 3 \delta^6)z_3
\nonumber\\
&&\qquad+\,
16\delta^2(1152 + 1024 \delta^2 - 140 \delta^4 - 53 \delta^6 - 2\delta^8)z_3^2
\nonumber\\ 
&&\qquad-\,8\delta^2(5376+128 \delta^2 - 1576 \delta^4-240\delta^6 - \delta^8)
z_3^3\nonumber\\
&&\qquad+\,2 (2048 - 768 \delta^2 - 19968 \delta^4 - 6968 \delta^6
 - 350 \delta^8 - 3 \delta^{10} )z_3^4\nonumber\\
&&\qquad-\,4(4096 - 20096 \delta^2 - 11168 \delta^4 - 1208 \delta^6 - 
43 \delta^8)z_3^5\nonumber\\
&&\qquad+\,(38912 - 75392 \delta^2 - 16960 \delta^4 - 996 \delta^6 
- 3 \delta^8)z_3^6\nonumber\\
&&\qquad-\,4(13312 - 6304 \delta^2 - 872\delta^4 - 3\delta^6)z_3^7 +
8(4512 - 500 \delta^2 - 15 \delta^4)z_3^8\nonumber\\
&&\qquad-\,\left.32(336-\delta^2)z_3^9 + 1280 z_3^{10}\right]\\
&&\quad +\,\left[8\delta^4(16+2\delta^2 + 3 \delta^4) - 
32\delta^4(8 - \delta^2)z_3\right.\nonumber\\
&&\qquad -\,2\delta^2(320 - 272 \delta^2 + 64 \delta^4 - 3 \delta^6)z_3^2 +
16\delta^2 (40 - 54\delta^2 - 5 \delta^4)z_3^3\nonumber\\
&&\qquad -\,(1024 - 720 \delta^4 - 3 \delta^6)z_3^4 +
8(96 - 36 \delta^2 - 5 \delta^4)z_3^5\nonumber\\
&&\left.\qquad+\,8(80+7\delta^2)z_3^6 - 512 z_3^7\right]\nonumber\\
&&\left.\qquad\times\delta^2(z_3^2 - 2)^4\ln
\frac{z_3\,\sqrt{4 + \delta^2 - 4 z_3} + 2\sqrt{(1-z_3)(z_3^2 - \delta^2)}}
{z_3\,\sqrt{4 + \delta^2 - 4 z_3} - 2\sqrt{(1-z_3)(z_3^2 - \delta^2)}}
\right\}\times\frac{1}{S_{\mathrm quark}(z_3)}\nonumber
\end{eqnarray}
\end{mathletters}

% ==========================================================================
% Figure captions
% ==========================================================================

% ==========================================================================
% References
% ==========================================================================


\begin{references}

\bibitem{Chang} C.H. Chang, Nucl. Phys. {\bf B172} 425 (1980).
\bibitem{Berger}E.L. Berger and D. Jones, Phys. Rev. {\bf D23} 1521 (1981).
\bibitem{Kuhn} J. H. K\"uhn, J. Kaplan and E. G. O. Safiani, Nucl. Phys. {\bf
 B157} 125 (1979).
\bibitem{Guberina} B. Guberina, J.H. K\"uhn, R.D. Peccei and R. R\"uckl, Nucl. 
 Phys. {\bf B174} 317 (1980).
\bibitem{Baier} R. Baier and R. R\"uckl, Z. Phys. C {\bf 19} 251 (1983).
\bibitem{Psidata} The CDF collaboration, Fermilab-Conf-94/136-E (1994), 
 unpublished.
\bibitem{Papadimitriou} The CDF collaboration, Fermilab-conf-95/128-E (1995), 
 unpublished.
\bibitem{Upsilondata} The CDF collaboration, Fermilab-Pub-95/271-E (1995), 
 unpublished.
\bibitem{Bodwin} G.T. Bodwin, E. Braaten and G.P. Lepage, Phys. Rev. {\bf D51} 
1125 (1995).
\bibitem{BraatenFleming} E. Braaten and S. Fleming, Phys. Rev. Lett. {\bf 74} 
 (1995) 3327.
\bibitem{CGMP}M. Cacciari, M. Greco, M.L. Mangano and A. Petrelli, 
 Phys. Lett. {\bf B356} 553 (1995).
\bibitem{ChoLeibov1} P. Cho and A.K. Leibovich, Phys. Rev. {\bf D53} 
150 (1996).
\bibitem{ChoLeibov2} P. Cho and A.K. Leibovich, Caltech preprint CALT-68-2026 
(hep-ph/9511315).
\bibitem{BraatenChen} E. Braaten and Y.-Q. Chen, Phys. Rev. Lett. {\bf 76} 
 730 (1996).
\bibitem{Kuhn1} J.H. K\"uhn and H. Schneider, Phys. Rev. {\bf D24} 
 2996 (1981).
\bibitem{Kuhn2} J.H. K\"uhn and H. Schneider, Z. Phys. {\bf C11} 
 253 (1981). 
\bibitem{Driesen} V.M. Driesen, J.H. K\"uhn and E. Mirkes, Phys. Rev. {\bf 
 D49} 3197 (1994).
\bibitem{Clavelli} L. Clavelli, Phys. Rev. {\bf 
 D26} 1610 (1982).
\bibitem{BCY} E. Braaten, K. Cheung and T.C. Yuan, Phys. Rev. {\bf D48} 
 4230 (1993).
\bibitem{CLEO1} The CLEO Collaboration, Phys. Rev. {\bf D50} 
 43 (1994).
\bibitem{CLEO2} The CLEO Collaboration, Phys. Rev. {\bf D52} 
 2661 (1995).
\bibitem{Wolf} A. Wolf, Private Communication.


\end{references}
\end{document}